\documentclass[10pt]{iopart}
\usepackage{graphicx}

\begin{document}
 \title{Longitudinal magnetoresistance in YBa$_2$Cu$_3$O$_7$ at high magnetic fields up to 100~T}
 \author{Shiyue~Peng$^1$, Xu-Guang~Zhou$^1$, Yasuhiro~H.~Matsuda$^1$, Qian Chen$^1$, Masashi Tokunaga$^1$, Yuto Ishii$^1$, Satoshi Awaji$^2$, Taiga Kato$^3$, Tomonori Arita$^3$, and Yutaka Yoshida$^3$}

\address{$^1$ Institute for Solid State Physics, University of Tokyo, Kashiwa, Chiba 277-8581, Japan}
 \address{$^2$ Institute for Materials Research, Tohoku University, Sendai, Miyagi 980-8577, Japan}
 \address{$^3$ Department of Electrical Engineering, Nagoya University, Nagoya, Aichi 464-8603, Japan}
\eads{\mailto{zhou@issp.u-tokyo.ac.jp}
\mailto{ymatsuda@issp.u-tokyo.ac.jp}}
 
 \begin{abstract}
 The investigation of transport characteristics in high-temperature superconductors under magnetic fields is one of the most important topics in condensed matter physics and transport applications. 
For YBa$_2$Cu$_3$O$_7$ (YBCO),
the measurements of magnetoresistance under a high magnetic field are technically challenging because the required magnetic field ($B$) to suppress the superconductivity is 100~T class.
The low temperature (from 52 to 150~K) longitudinal magnetoresistance ($B$$\parallel$ab-plane$\parallel$$J$, where $J$ is an electrical current) was measured up to 103~T in optimally doped YBCO thin films.
A radio frequency reflection method and the single-turn coil technique were employed.
The electrical resistivity $\rho_{ab}$ exhibited a non-saturating magnetoresistance behavior until the highest field region,
with the slope $\beta (=d \rho/d B)$ showing a pronounced deviation compared to the transverse magnetoresistance ($B$$\parallel$c-axis) case. 
These findings suggest a potential contribution due to the non-orbital origin in the high field phase of YBCO,
because the quasiparticle orbital motion is expected to be absent in the longitudinal magnetoresistance.

 \end{abstract}
\vspace{2pc}
\noindent{\it Keywords\/}: Electrical properties, Magnetic fields, High temperature superconductors, RF waves, Longitudinal magnetoresistance

\maketitle
\ioptwocol
 \begin{figure*}
 \centering
 \includegraphics[width = 0.9\linewidth]{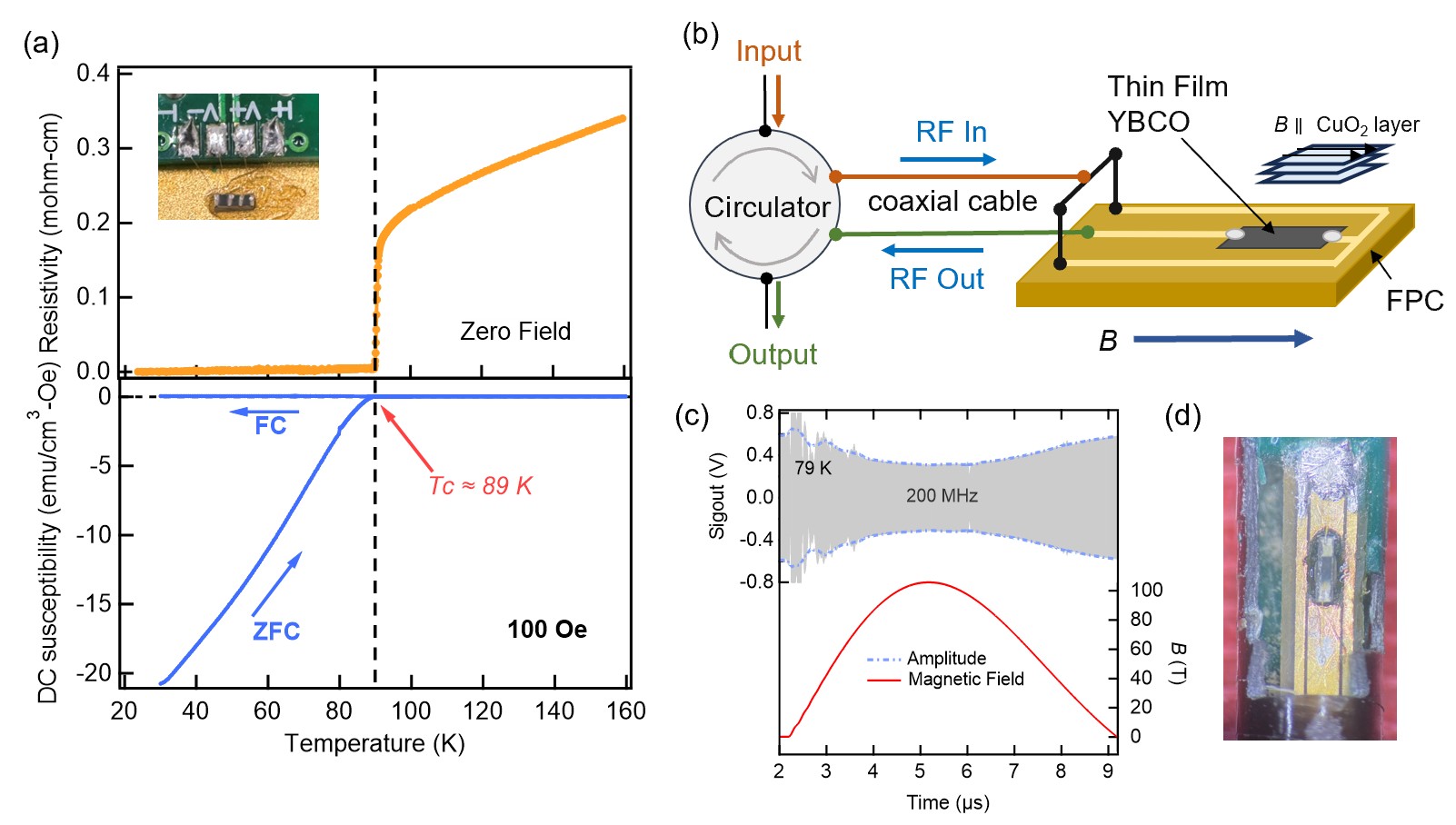}
 \caption{(a) The electrical resistivity and DC magnetic susceptibility versus temperature measurements for confirming the YBCO thin film sample quality and $T_c$.
 (b) The schematic diagram of the RF (Radio Frequency) measurement system.  
 (c) The typical raw data of the RF reflection signal at 79~K with its envelope curves together with the magnetic field waveform.
 (d) A photo of the end of FPC connected with a thin film YBCO sample.}
\label{RF}
\end{figure*}

\section{Introduction}
The high-temperature cuprate superconductors are quite intriguing topic in the condensed matter physics because it is not only a typical quantum many-body problem~\cite{keimer2015, Sachdev2003}, 
but also a problem leading to potential electrical power transport applications such as superconducting wires~\cite{Scanlan2004,Noe2007, larbalestier2001} or magnets~\cite{hahn2019} for their high critical temperature.
It is noted that their resistivity $\rho_{ab}$ near optimal doping grows linearly with temperature over a broad temperature range~\cite{Gurvitch1987, Anderson1992, Hussey2008},
which is a hallmark of strange metal who lies outside the standard Fermi-liquid description of metals~\cite{Anderson1988,Ioffe1989,phillips2022, Kapitulnik2019, yuan2022, Sarkar2019}.
This behavior is also observed in heavy-fermion compounds, pnictides, and organic superconductors~\cite{hayes2016,doiron2009,WEICKERT2006,analytis2014}.
This $T$-linear resistivity is often associated with Planckian dissipation~\cite{hartnoll2022}, 
in which the scattering time $\tau\sim\hbar/k_BT$, 
where $\hbar$ is the reduced Planck constant, $k_B$ is the Boltzmann constant.
This condition is believed to arise in a strongly interacting critical state, rooted at a quantum critical point (QCP)~\cite{hartnoll2022,phillips2022}, 
where a critical end point of the phase transition is driven to zero temperature.

Similarly, when superconductivity of cuprates is suppressed by magnetic fields,
the dependence of resistivity on magnetic field $B$ is expected to be another facet of Planckian dissipation in strange metals as reported in~\cite{giraldo2018, Ayres2021}.
The electrical resistivity would have not only exhibited linear-in-temperature behaviour, but also an anomalous linear-in-field and non-saturating magnetoresistance (MR) behavior in the strange metal phase. 
The underlying cause of this phenomenon of MR remains a subject of debate,
several theoretical explanations have been proposed,
involving multiband effect~\cite{xing2016,Xing2018,Ahmad2024}, the presence of disorder~\cite{Patel2018,Singleton2020} 
and orbital motion of electrons around the Fermi surface with anisotropic impurity scattering~\cite{Hinlopen2022,ataei2022}.

Although it has been suggested that the origin of this non-saturating linear MR differs from that of the T-linear behavior in strange metal and may not be dominated by quantum criticality~\cite{xing2016,Xing2018,Ahmad2024,Patel2018,Singleton2020,
Hinlopen2022,
ataei2022}, 
as it can be explained within the quasiparticle framework.
It is important to note that most of the previous experiments of cuprates have been limited to out-of-plane magnetic field directions, 
specifically $B$$\parallel$c-axis,
where the plane denotes the CuO$_2$ plane.
They correspond to the transverse magnetoresistance (TMR) because the electrical current $J$ is also in-plane ($B$$\perp$$J$).
For the in-plane magnetic field direction with in-plane resistance, i.e., $B$$\parallel$ab-plane$\parallel$$J$ (longitudinal magnetoresistance, LMR),
 theoretically no cyclotron motion and related scattering are expected within the classical framework. 
This means that if the similar non-saturating magnetoresistance phenomenon occurs under in-plane field LMR, 
as reported in~\cite{Ayres2021} 
which observed large LMR in cuprates Tl2201 and Bi2201,
the non-orbital origin possibly contributes to the MR.
Also, the influence of vortexes is diminished in this direction, resulting in mechanisms that require consideration being more distinct.

Here, in this study, 
we investigated the LMR behavior under the in-plane magnetic field. 
We used YBa$_2$Cu$_3$O$_7$ thin films near optimal doping,
which is one of the cleanest and ordered cuprates, 
to explore the possible anomalous properties of MR.
The LMR in such high-$T_c$ YBCO at low temperatures and high magnetic fields has never been well investigated because 
ultra-high magnetic fields exceeding 100~T are required to suppress the superconducting phase~\cite{miura2002,sekitani2003}.
Generating such high magnetic fields can only be achieved through a pulsed destructive manner~\cite{Miura2001STC,nakamura2018}. 
It is challenging to investigate the electric resistivity properties in such a system due to the extremely short duration time and electromagnetic noise. 
Here, we accomplished this experiment by developing a radio frequency (RF) reflection method adapted to the single-turn coil magnetic field generator in the Institute for Solid State Physics. 
LMR curves were obtained over a temperature range spanning from 52 to 150 K.
Once the superconducting phase is suppressed, the high field phase demonstrates a non-saturating MR in the high magnetic field region, suggesting a possible appearance of the strange metal phase. 
In comparing the slopes of MR $\beta (=d \rho/d B)$ of magnetoresistance of YBCO between $B$$\parallel$c-axis$\perp$$J$,
and $B$$\parallel$ab-plane$\parallel$$J$ cases,
the difference lets us consider the potential contributions of a non-orbital origin of MR. 
The quantum effect from strange metal features may play a role in dominating the MR behavior in YBCO, 
suggesting the anomalous in such a high-field phase.

\section{Experiment}

The experiment in this work employs an optimally-doped YBa$_2$Cu$_3$O$_7$ thin film sample with a critical temperature $T_c$ of approximately 89~K, as illustrated in Fig.~\ref{RF}(a) characterized by the measurements of electrical resistivity and DC susceptibility versus temperature in Physical Property Measurement System (PPMS) and Magnetic Property Measurement System (MPMS), respectively.
The film, with a thickness of 200~nm, is deposited on a SrTiO$_3$ (100) substrate. 
To ensure good electrical connectivity, silver is sputtered onto both ends of the sample's surface, connecting it to thin Au wires via silver paste.
Pulsed magnetic fields up to 103~T are generated using a destructive method, specifically the vertical-type single-turn coil technique~\cite{Miura2001STC}. 
The magnetic field is applied parallel to the 
ab-plane and is also parallel to the thin film, which is (001) oriented, as shown in Fig.~\ref{RF}(b). 
For impedance measurement, a radio frequency (RF) reflection system is developed.

In Fig~\ref{RF}(b), the RF method is implemented using a circulator, with input, output, and sample ports. 
The frequency of the RF incident wave is set at 200~MHz (input), 
which is over two orders of magnitude higher than the pulsed magnetic field waveform frequency (approximately 200~kHz).
For the sample port, a 150 mm-long flexible printed circuit (FPC) with a fabricated three-conductor coplanar transmission line (CTL) is used in place of a coaxial cable. 
This choice aims to prevent mechanical breakage of the coaxial cable when subjected to an intense, fast-pulsed magnetic field. 
This breakage occurs due to a strong inward force on the outer conductor caused by a substantial induction current interacting with the magnetic field.
The FPC impedance is designed to be 50~$\Omega$ to ensure optimal impedance matching for the high-frequency circuit. 
The central line of the CTL has a hole to hold the substrate of the YBCO thin film, aligned in the $B$$\parallel$
ab-plane$\parallel$$J$ direction. 
Fig.~\ref{RF}(d) presents a photograph of the setup around the sample. 
The CTL transmits the RF signal (RF~in) to the sample and carries the reflected RF signal from the sample (RF~out).
The reflection signal is a feature of the sample's impedance. 
Consequently, MR can be determined by detecting the magnetic field dependence of the RF signal reflection (output). 
Similar methods are also discussed in Ref.~\cite{kane1997,shitaokoshi2023}.

 \begin{figure}
    \includegraphics[width = .8\linewidth]{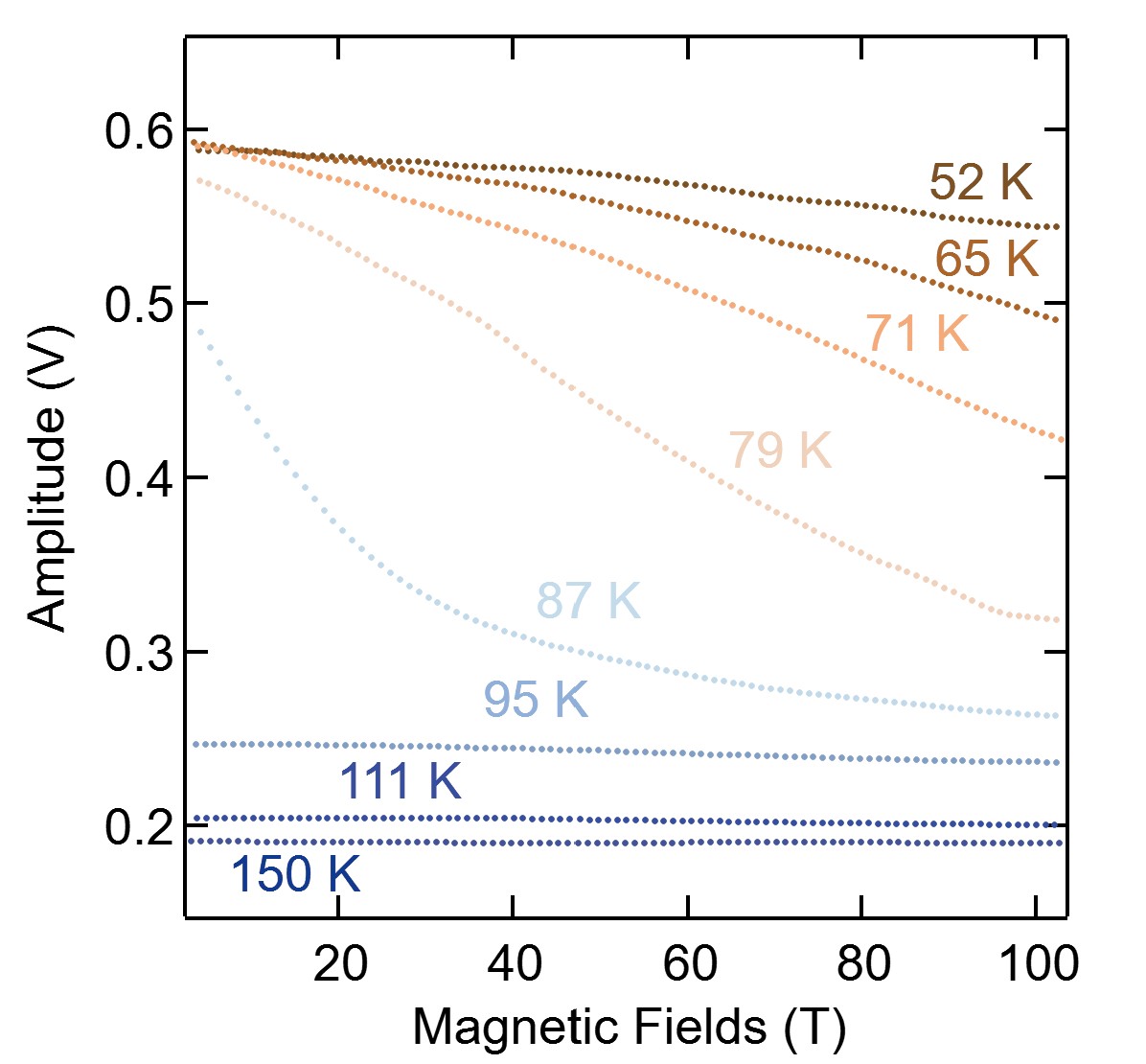}
     \caption{The magnetic field dependence of the radio frequency reflection amplitude of YBCO for temperatures ranging from 52 K to 150 K up to 103~T.}
     \label{RD}
    \end{figure}

The magnetic field components perpendicular to the FPC-plane ($B_{\perp}$) generated by the single-turn coil technique cannot be neglected~\cite{Nakao1985}, leading to induced electromagnetic noise. 
Consequently, the position of the FPC is carefully adjusted geometrically to minimize the induced voltage 
which is proportional to $\partial{B}_{\perp}/\partial t$.
The three-line CTL offers an advantage in that the voltage measured between the inner and outer conductors is designed not to be affected by $\partial{B}_{\perp}/\partial t$. 
This is achieved because the two closed circuits, consisting of the center and two side lines, generate voltages with opposite signs, allowing cancellation of $\partial{B}_{\perp}/\partial t$ effects~\cite{kane1997}.
The RF output signal is transmitted by a thin but semi-regid coaxial cable connected to the edge of the FPC, where the magnetic field is nearly negligible. 
To control the temperature from 50 K to 150 K, a liquid helium flow cryostat with a tail part made of plastic is utilized, and the temperature is managed by controlling the flux flow rate~\cite{Matsuda2018}.
Under this experimental setup, the temperature in the sample region remains stable, 
and the heating effect of the thin film sample during the measurement can be negligible. (See Fig.~S4 and Fig.~S5 in Supplementary Materials for details.)

\section{Results and Discussion}

The raw data of the RF signal reflection at 79 K with its envelope curves are shown as a function of time in Fig.~\ref{RF}(c) together with the magnetic field waveform.
The duration time of the magnetic field is approximately 6 ${\mu}$s, and the obtained maximum field is 103~T.
The gray area represents the obtained output signal.
The blue lines in Fig.~\ref{RF}(c) represent the envelope curves that show the amplitude of the reflected signal which is extracted from the numerical lock-in technique.
At the onset of magnetic field generation,
a significant switching electromagnetic noise is unavoidably produced due to the injection of mega-ampere driving electric currents.
Consequently, our analysis focuses solely on the data during the field-decreasing process.

By correlating the time scale of the magnetic field $B$($t$) and the amplitude $A$($t$),
we obtain the magnetic field dependence of the RF signal amplitude $A$($B$), which is shown in Fig.~\ref{RD}.
The $A$($B$) has been measured at varying temperatures (150, 111, 95, 87, 79, 71, 65, and 52 K) up to 103~T. 
The signal-to-noise (S/N) ratio is improved compared with the previous results~\cite{Brien2000,miura2002,sekitani2003}, 
in which the transport properties are also investigated with the destructive manner of ultra-high magnetic field generation.

The temperature dependence of the electrical resistance $R$($T$) as well as the temperature dependence of the RF amplitude $A$($T$) is measured to make the calibration curves $R$($A$).
This measurement allows us to experimentally establish a relationship between $A$ and $R$, rather than obtaining it by analyzing the phase with an admittance chart~\cite{shitaokoshi2023}.
The calibration curves $R$($A$) are renewed before and after each explosion of the single-turn coil experiment~\cite{Miura2001STC}, as shown in Fig.~S2.
This is to reduce the effect of impedance change of the setup caused by the explosion.
Thus, by mapping the temperature dependence of $A$ and $R$, 
we finally obtain the LMR curves $R$($B$) of YBCO via the corresponding transformation from the $A$($B$) in Fig.~\ref{RD}.

    \begin{figure}
    \includegraphics[width = .8\linewidth]{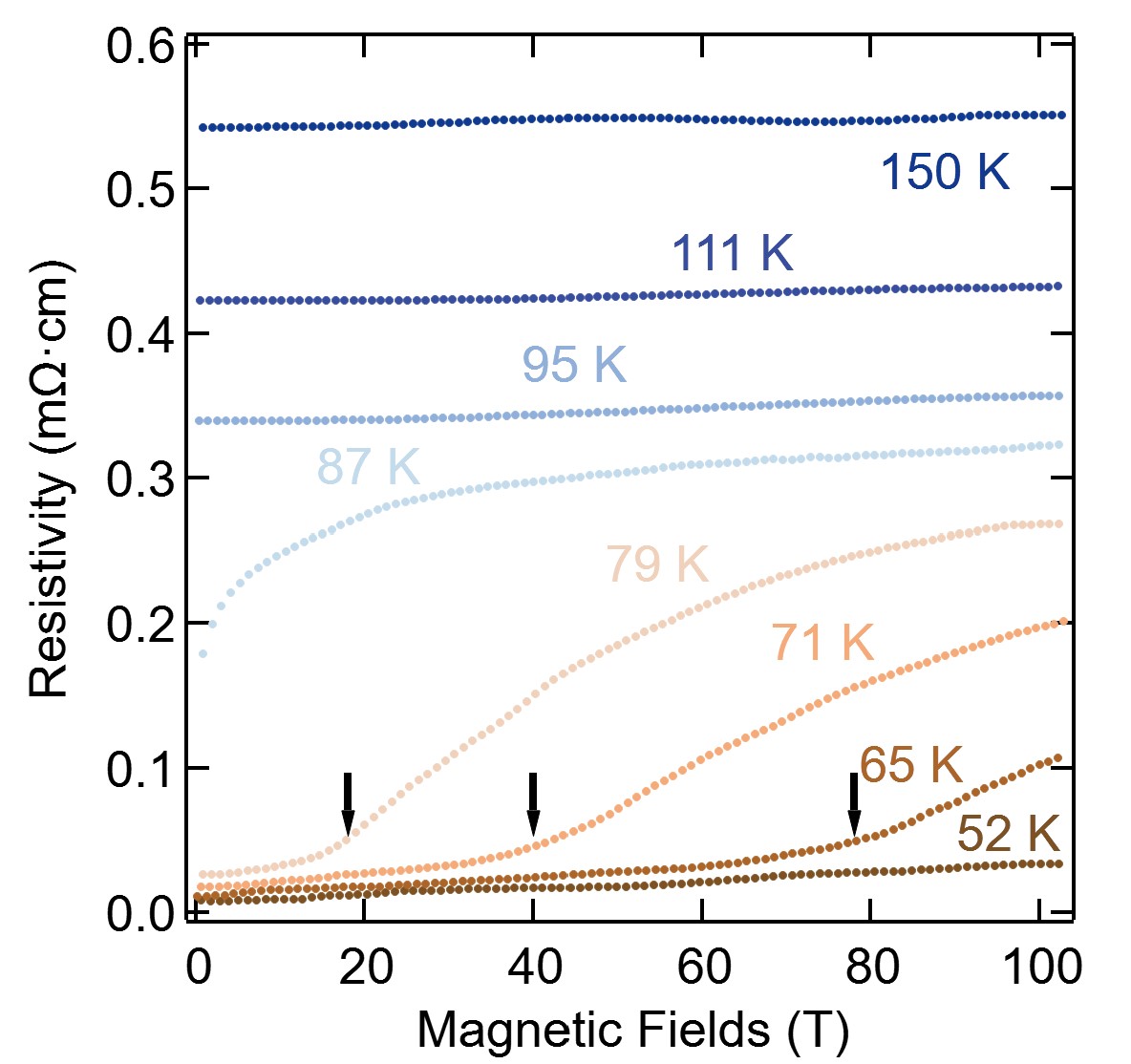}
     \caption{LMR of YBCO for temperatures ranging from 52 K to 150 K in field of up to 103~T. The arrows indicate the magnetic field positions of the kink structure where the superconductivity starts to be suppressed.}
     \label{MR}
    \end{figure}

The LMR curves are exhibited in Fig.~\ref{MR}. 
At 52 K, 
only signs of the phase transition were observed.
At 65, 71, 79, and 87~K, an increase of the resistivity is observed with increasing magnetic field. 
The kink-structure is observed under certain magnetic fields of 18, 40 and 78~T at 65, 71, and 79~K 
as indicated by the black arrows, respectively, 
which is defined as the onset of the suppression of superconductivity here following Mackenz \textit{et al.}~\cite{mackenzie1993}.
At temperatures above $T_{\rm{c}}$ (95, 111, and 150~K) 
and the high field phase region of 87~K, 
the LMR is likely exhibiting a non-saturating behavior concerning $B$ with a smaller slope.
As the temperature increases, the magnetic field position where the kink in LMR appears gradually shifts to lower magnetic fields, disappearing at 95~K when the temperature exceeds $T_{\rm{c}}$ ($\approx$ 90~K). 
The upper critical magnetic field ($B_{c2}$) is defined as the magnetic field where the resistivity reaches 90$\%$ of the resistivity of the normal phase near the superconducting-normal phase transition. This definition aligns closely with the approach outlined in Ref.~\cite{mackenzie1993, Brien2000}.
The obtained $B_{c2}$ are closely aligned with the phase boundary previously determined using the transmission method in pulsed magnetic field experiments~\cite{sekitani2007,sekitani2004} (See in Fig.~S3), 
demonstrating that our results are reliable.

\begin{figure}
    \includegraphics[width = 1.0\linewidth]{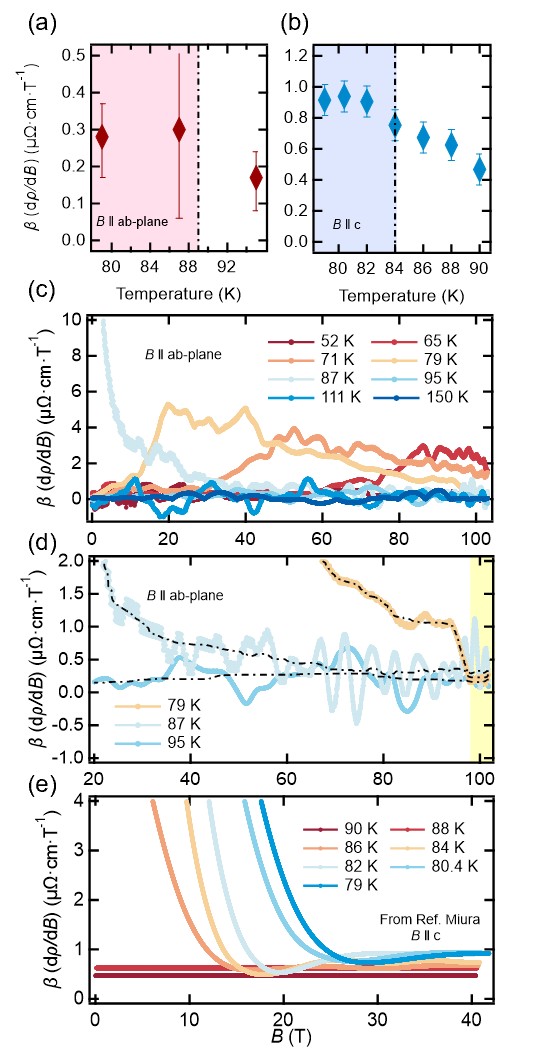}
   \caption{Temperature dependence of average $\beta$ ($= d\rho/dB$) at
the highest measured field range (the last 5~T) for the MR curves obtained in (a) $B$$\parallel$ab-plane$\parallel$$J$ done in this work at 79, 87 and 95~K. 
     (b) at the $B$$\parallel$c-axis directions from Ref.~\cite{miura2002}. The error bars reflect the uncertainty in the $\beta$ values. The color-shaded area represents the superconducting region of samples with critical temperature $T_c$ indicated by the dashed line. 
     (c) The B-dependence of $\beta$ values across the entire magnetic field range obtained from the LMR in this study. (d) The detailed average $\beta$ values estimation at 79, 87 and 95~K. Estimates for $\beta$ for which the non-superconducting state was not fully reached (71, 65 ,52~K) and significant uncertainty exists (111 and 150~K) are not presented. The yellow region represents the last 5~T in each sweep and the dashed lines shows the average of $\beta$.(e) The B-dependence of $\beta$ values obtained form TMR in ref.~\cite{miura2002} using a non-destructive magnetic field system.}
     \label{beta}
    \end{figure}

In Fig.~\ref{MR}, the LMR at 95 and 111~K exhibits a non-saturating behavior up to 103~T. 
A similar behavior is also observed at 87~K over the high magnetic field range after entering the high field phase.
While at 71~K and 65~K, 
the LMR curves show that the phase transition process continues until the highest magnetic field.
To discuss the LMR behavior in more detail, we introduce the field slope parameter $\beta$ ($= d\rho/dB$) by taking the differential of the LMR curves as shown in Fig.~\ref{beta}(c).
At 79~K, 
the field dependence of slope $\beta$ reveals a stabilization trend at high magnetic field region, leading us to believe that the phase transition has been completed by this point. 
For 87~K, 
it exhibits a typical differential feature that has undergone the complete superconducting-normal phase transition and ultimately exhibits the resistivity characteristics of the high field phase.
The slope $\beta$ of LMR at 95~K does not exhibit any quadratic feature as reported in many other researches~\cite{ataei2022,Hinlopen2022,Ayres2021,Sarkar2019,giraldo2018}, but instead maintains an almost constant $\beta$ value up to the highest magnetic field.
A summary of $\beta$ values in the high field phase is presented in Fig.~\ref{beta}(a),
where the temperature dependence of $\beta$ at the highest measured field range (over the last 5~T of each sweep, where the high field phase can be accessed) is plotted.
Although there is uncertainty in the $\beta$ values we obtained, 
by combining the trend of the LMR curves in Fig.~\ref{MR} and calculating the average of differential curves (as shown in Fig.~\ref{beta}(d)), we can still estimate $\beta$ and believe they possess finite values.
The striking feature is that, 
our magnetic field direction is parallel to the applied current in CuO$_2$-plane.
Given the quasi-two-dimensional layered structure of YBCO,
the orbital motion of the quasiparticles is not expected in this case.
But there still exhibits non-saturating and approximately linear LMR in our results. 
Even if there is some warping on the Fermi surface cylinder or experimental error in sample setting, 
according to the Boltzmann theory, it is generally expected that only a minimal or even negligible magnetoresistance should appear under this configuration~\cite{ataei2022}.

To compare whether the magnitude of LMR we obtained is sufficiently small,
the same analysis is also conducted for the TMR results reported by Miura \textit{et al.}~\cite{miura2002} in $B$$\parallel$c-axis as shown in Fig.~\ref{beta}(e),
with which the superconductivity is more easily suppressed.
The experiment done by Miura \textit{et al.} was performed using a YBCO thin film under $B$$\parallel$c-axis up to 50~T. 
They also exhibit an almost linear and non-saturating behavior in the high field phase.
The $\beta$ values obtained from the TMR curves are plotted as shown in Fig.~\ref{beta}(b).
Noting that the $T_c$ ($\approx$ 84~K) of the YBCO thin film in~\cite{miura2002} is slightly lower than the thin film used in our work.
The $\beta$ is found to increase with decreasing temperature and to saturate at a constant value below $T_c$,
which is consistent with previous measurements in~\cite{giraldo2018,ayres2024} on other cuprates.
It should be pointed out that $\beta$ in Fig.~\ref{beta}(a) for $B$$\parallel$ab-plane$\parallel$$J$ (LMR) is roughly half of $\beta$ in Fig.~\ref{beta}(b) for $B$$\parallel$c-axis$\perp$$J$ (TMR).
Unlike the findings in~\cite{Ayres2021} on heavily
overdoped Bi$_2$Sr$_2$CuO$_{6+\delta}$ 
and Tl$_2$Ba$_2$CuO$_{6+\delta}$, where the slopes of MR in two directions are comparable,
the distinct difference in $\beta$ for YBCO is very pronounced. 
On the other hand,
this difference is not as pronounced as in~\cite{ataei2022} on overdoped La$_{1.6-x}$Nd$_{0.4}$Sr$_x$CuO$_4$,
where the MR can be considered negligible under the $B$$\parallel$ab-plane$\parallel$$J$ (LMR) conditions.
Although there is significant uncertainty, our estimated $\beta$ in Fig.~\ref{beta}(a) does indicate a finite value on average.
This suggests that, for our results, the mechanisms behind the non-saturating MR may include non-classical magnetic field effects.
Also we note that the finite LMR observed in our YBCO sample is smaller than that reported in Bi$_2$Sr$_2$CuO$_{6+\delta}$ 
and Tl$_2$Ba$_2$CuO$_{6+\delta}$~\cite{Ayres2021}, 
which may arise from differences in doping level and electronic structure. 
While their samples are heavily overdoped single-layer cuprates, ours is near optimal doping and has bilayer structure with CuO chain contributions. 
These factors can affect scattering processes and reduce the field sensitivity of transport. 
Nevertheless, the overall magnitude of finite LMR remains comparable between our results ($\sim$0.3
$\mu \Omega \cdot \mathrm{~cm} \cdot \mathrm{~T}^{-1}$) and previous study ($\sim$0.35$\mu \Omega \cdot \mathrm{~cm} \cdot \mathrm{~T}^{-1}$ in Tl$_2$Ba$_2$CuO$_{6+\delta}$~\cite{Ayres2021}), suggesting a possibly common origin across different compounds of cuprates.

The Boltzmann transport theory within the framework of Fermi liquid quasiparticles can explain part of the MR as also has been confirmed by~\cite{ataei2022,Hinlopen2022,Ayres2021} in the TMR ($B$$\perp$$J$) cases,
which mainly depends on the orbital motion of quasiparticles in the presence of anisotropic impurity scattering. 
In the case of the LMR ($B$$\parallel$$J$),
the absence of conventional cyclotron motion and related scattering is expected;
thus, the finite positive LMR observed in our experiment up to 103~T cannot be interpreted by the above-mentioned orbital origin. 
We propose other potential contributions exist that are related to the non-orbital origin.
In this case,
the magnetoresistance is
considered to originate from the influence of the
magnetic field on the dynamics of the critical fluctuations in the quantum critical region that govern
the relaxation time~\cite{zaanen2004,Aji2007,Lederer2017}.
Consequently, the $B$-linear behavior of the LMR we
observe in this work can be one of the hallmarks of the strange metal phase near the quantum critical
point.

Additionally,
it has been proposed recently that electronic transport in cuprates may involve two contribution channels~\cite{Ayres2021,ayres2024,hayes2021,Licciardello2019}. 
On the one hand, the quasiparticle component preserves the momentum information and the orbital character of the Fermi surface, leading to angle-dependent scattering.
On the other hand, 
there exists a class of isotropic, non-quasiparticle carriers that no longer undergo orbital motion, 
yet dominate the magnetoresistance and $T$-linear resistivity. The scattering rate of these carriers is constrained by the Planckian
limit. 
This class of Planckian dissipators does not exhibit an orbital response to magnetic fields. 
As shown in Fig.~\ref{beta}, 
although the strength of magnetoresistance in YBCO remains of comparable magnitude with different field orientations, the slope $\beta$ exhibits a noticeable degree of anisotropy. 
In the configuration of TMR, 
quasiparticle orbital motion can definitely contribute to the magnetoresistance as discussed previously, 
which is consistent with the observed enhancement of the slope $\beta$ ($\sim$0.9$\mu \Omega \cdot \mathrm{~cm} \cdot \mathrm{~T}^{-1}$) compared to the LMR configuration ($\sim$0.3$\mu \Omega \cdot \mathrm{~cm} \cdot \mathrm{~T}^{-1}$) in our measurements when the orbital effect is absent.
Our findings that the persistence of $B$-linear and non-saturating magnetoresistance in a field orientation that suppresses orbital contributions can suggests that a non-quasiparticle transport channel,
possibly governed by Planckian dissipation,
should play an important role.
Given that our YBCO samples are optimally doped and lie near the quantum critical point, 
such a non-orbital origin should be rooted in the quantum criticality of the strange metal phase.
And whether the high field $B$-linear magnetoresistance observed in TMR of cuprates is governed purely by orbital effects or arises from a combination of orbital and strange metal related non-orbital contributions remains a question that requires further investigation.
Here, we also should note that the possible misalignment of the magnetic field direction in our experiments is expected to be smaller than $1^\circ$, which, according to Hinlopen et al.~\cite{Hinlopen2022}, can result in a finite magnetoresistance due to the orbital effect as large as 0.01$\%$. This is safely smaller than the LMR found in this study.
In summary,
The non-saturating and nearly $B$-linear LMR observed in our experiment may not be conventional, 
but is related to the nature of the strange metal phase in cuprates at ultrahigh magnetic field.

Here, we note that the magnetic field of 100~T is not sufficient to break superconductivity at lower temperatures. According to~\cite{sekitani2007},
the upper critical magnetic field at 4.2 K in our experiment direction 
($B$$\parallel$CuO$_2$-plane) is about 250~T.
The LMR at lower temperatures far below $T_c$ is therefore a promising future issue to be clarified. 
Electromagnetic flux compression~\cite{Cnare1966, Novac2004} is required to generate an ultrahigh magnetic field in the range of 1000~T~\cite{nakamura2018} that would be necessary for the future study.

\section{Conclusion}
In summary,
we have measured the LMR in YBa$_2$Cu$_3$O$_7$ at varying temperatures via the radio frequency reflection method in ultrahigh magnetic fields exceeding 100~T with the condition of $B$$\parallel$CuO$_2$-plane$\parallel$$J$.
For the non-saturating and nearly $B$-linear LMR observed in the high field phase,
we analyzed the slope parameter $\beta$ at the highest magnetic field region, 
observing that its trend is consistent with other cuprates in previous research, 
where a saturation value appears at low temperature. 
Given that the LMR behavior observed in our experiment cannot be explained by orbital motion in a quasiparticle framework like in the TMR case~\cite{Hinlopen2022,ataei2022}, 
we propose that the contribution from a non-orbital origin, 
associated with the quantum criticality in the strange metal phase, 
potentially contributes to the LMR of YBCO.

\ack{S.P. and X.-G.Z. thank for the fruitful discussions and experimental technique supports from Takashi Shitaokoshi, Yoshimitsu Kohama, Kazuki Matsui, Fengfeng Song, Akira Matsuo and Wei Li. This work was funded by the JSPS KAKENHI, Grant-in-Aid for Transformative Research Areas (A) Nos.23H04859 and 23H04860, Grant-in-Aid for Scientific Research (B) No. 23H01117. X.-G.Z. and Y.H.M was funded by JSPS KAKENHI No. 22F22332.}
\section*{References}

\bibliographystyle{iopart-num}

\providecommand{\newblock}{}

\end{document}